\documentclass[structabstract]{aa} 
\usepackage{longtable}
\usepackage[tableposition=below]{caption}
\usepackage{tablefootnote}
\usepackage{lscape}
\usepackage{graphicx}
\usepackage{epstopdf}
\usepackage{txfonts}
\usepackage{hyperref}

\begin{document} 

\title{Hydrodynamic simulations of the recurrent nova T Coronae Borealis: Nucleosynthesis predictions}

\author{Jordi Jos\'e \inst{1,2}
   \and Margarita Hernanz \inst{2,3}
    }

\offprints{J. Jos\'e}

 \institute{Departament de F\'\i sica, EEBE,
            Universitat Polit\`ecnica de Catalunya (UPC),
            c/Eduard Maristany 16,
            E-08019 Barcelona,
            Spain\
        \and
            Institut d'Estudis Espacials de Catalunya (IEEC),
            c/Esteve Terradas 1,
            E-08860 Castelldefels,
            Spain\
        \and
            Institut de Ci\`encies de l'Espai (ICE-CSIC),
            Campus UAB, 
            Cam\'\i \, de Can Magrans s/n, 
            E-08193 Bellaterra,
            Spain
            \\
      \email{jordi.jose@upc.edu}}

\date{\today}

\abstract{Recurrent novae are, by definition, novae observed in outburst more than once or identified by the presence of vast super-shells,
ejected in previous eruptions, surrounding the system. These systems are characterized by remarkably short recurrence times between outbursts,
typically ranging from
1 to about 100 yr. Such short recurrence times require very high mass-accretion rates, white dwarf masses approaching the Chandrasekhar limit,
and very high initial white dwarf luminosities.}
{T Coronae Borealis (T CrB) is one of the eleven known recurrent novae in our Galaxy. It was observed in outburst in 1866 and 1946, with
additional likely eruptions recorded in 1217 and 1787. Given its predicted recurrence period of approximately 80 yr, the next outburst is
anticipated to occur imminently, thus motivating a thorough examination of the main characteristics of this system.}
{We present 11 new hydrodynamic models of the explosion of T CrB for different combinations of parameters
(i.e., the mass, composition, and initial luminosity of the white dwarf, the metallicity of the accreted matter, and the mass-transfer rate).
We also report on 8 additional hydrodynamic models that include mixing at the interface between the accreted envelope and the outermost
layers of the underlying white dwarf,  and 3 models for 1.20 M$_\odot$ white dwarfs.}  
{We show that mass-accretion rates of $\dot M_{\rm acc} \sim 10^{-8} - 10^{-7}$ M$_\odot$ yr$^{-1}$ are required to trigger an outburst after 
80 yr of accretion of solar-composition material onto white dwarfs with masses M$_{\rm WD} \sim 1.30 - 1.38$ M$_\odot$ and initial luminosities 
L$_{\rm WD} \sim 0.01 - 1$ L$_\odot$.  For lower white dwarf luminosities, less massive white dwarfs, or reduced metallicity in the accreted material,
higher mass-accretion rates are required to drive an explosion within this timescale.
A decrease in metallicity or initial white dwarf luminosity leads to higher accumulated masses and ignition pressures,
resulting in more violent outbursts. These outbursts exhibit higher peak temperatures, higher ejected masses,
and greater kinetic energies.
Models computed for different white dwarf masses but identical initial luminosities reveal significant
differences in the elemental abundances of a wide range of species, including Ne, Na, Mg, Al, Si, P, S, Ar, K, Ca,
and Sc. These compositional differences offer a potential diagnostic tool for constraining the parameter
space and discriminating between the various T CrB models reported in this study.}
  {}

\keywords{(Stars:) novae, cataclysmic variables --- (Stars:) binaries: close --- Nuclear reactions, nucleosynthesis, abundances --- hydrodynamics}

\titlerunning{Hydrodynamic simulations of T CrB}
\authorrunning{Jos\'e \& Hernanz} 

\maketitle

\section{Introduction}
Classical novae are explosive stellar events that occur within close binary systems and are 
characterized by orbital periods spanning typically from $\sim 1.5$ to  15 h. 
In the canonical scenario (see Starrfield et al. 2008, 2012, 2016; Jos\'e \& Shore
2008; Jos\'e 2016; Chomiuk, Metzger \& Shen 2021 for reviews),
a white dwarf (WD) star is fed by a low-mass stellar companion, 
often a K-M main sequence star (although observations occasionally reveal the presence 
of more evolved companions). This mass transfer occurs as the secondary star
 overfills its Roche lobe, which allows matter outflows 
through the inner Lagrangian point of the system. The matter transferred (typically 
at a rate $\dot M_{\rm acc} \sim 10^{-8} - 10^{-10}$  M$_\odot$ yr$^{-1}$) forms an accretion
disk that orbits the WD star. A fraction of this hydrogen-rich disk 
spirals in, gradually accumulating on top of the WD
under semi-degenerate conditions. Compressional heating increases the temperature 
of the envelope until nuclear reactions set in, triggering a thermonuclear runaway 
(TNR). The TNR drives peak temperatures to around $(1 - 4) \times 10^8$ K and 
 leads to  the ejection of about $\sim 10^{-7} - 10^{-4}$ M$_\odot$ of nuclear-processed 
material at velocities of several thousand  km s$^{-1}$. In fact, the suite of nuclear processes that 
operate during the explosion result in non-solar isotopic abundance ratios within 
the ejected material (Gehrz et al. 1998; Downen et al. 2013; Kelly et al. 2013).

Novae are relatively common, ranking as the second-most-frequent 
type of stellar thermonuclear explosions in our Galaxy, following type I X-ray bursts. Although only a 
handful of novae are discovered per year, typically between five and ten, 
predictions based on Galactic and extragalactic data suggest a much higher occurrence rate 
of $\sim 50^{+31}_{-23}$ yr$^{-1}$ (Shafter 2017). This relatively high frequency is
partly due to the fact that neither the WD nor the binary system are disrupted 
during a nova explosion, and hence, nova outbursts are expected to recur. The predicted 
recurrence times for classical novae lie typically in the range $10^4 - 10^5$ yr. 
However, a 
subclass of novae, known as recurrent novae (RNe) -- by definition, novae observed 
in outburst more than once -- exhibit  recurrence periods ranging from as short
as 1 yr\footnote{Models predict  even shorter recurrence times, with some as brief as 50 days 
(Hillman et al. 2015), although no observational evidence of such systems has  been reported 
to date.} (e.g., M31N 2008-12a; Nishiyama \& Kabashima 2008, Darnley et al. 2016, and references 
therein) to about 100 yr (e.g., V2487 Oph; Schaefer 2010). Such notably short recurrence times 
require very  high mass-accretion rates, WD masses close to the Chandrasekhar 
limit (the mean WD mass of the ten RNe listed in Shara et al. 2018 is 1.31 M$_\odot$), 
and very high initial WD luminosities or temperatures (see Starrfield, 
Sparks \& Shaviv 1988, Yaron et al. 2005, Hernanz \& Jos\'e 2008).
Whether these recurrence times form a nearly continuous sequence, spanning from the short values 
observed in RNe to the longer values predicted for classical novae, remains a topic of debate.

Recurrent novae are symbiotic systems consisting of an accreting WD and a red giant companion in wider 
orbits (longer orbital periods) than classical novae.
Only  11 RNe have been discovered in the Milky Way\footnote{KT Eridani (Nova Eridani 2009) has been claimed 
to be the 11$^{\rm th}$ Galactic recurrent nova (Pagnotta \& Schaefer 2014, Schaefer et al. 2022, Shara et al. 2024), 
based on the discovery of a vast super-shell (50 pc in diameter), 
ejected in previous eruptions, centered around that object.}, although several others have been observed 
in the Andromeda Galaxy and the Large Magellanic Cloud. RNe exhibit a much broader range 
of orbital periods compared to classical novae, spanning from a few hours to several hundred days. 
Various classifications have been proposed based on different orbital periods, 
the presence of a plateau in the tail of the light curve, varying amplitudes 
and recurrence times, or distinct mechanisms driving the high mass-accretion 
rates inferred (see Schaefer 2010, and references therein). 
One well-studied RN is T Coronae Borealis (T CrB), a famous interacting binary system
observed in outburst in 1866 and 1946, and likely also in 1217 and 1787 (Schaefer 2023). 
With a predicted recurrence period of approximately 80 yr, its next outburst is expected to occur 
imminently, thus motivating a thorough examination of the main characteristics of this system. 
Specifically, we aim to investigate the conditions required to power an outburst with a 
periodicity of about 80 yr. Moreover, the interest in RNe as potential progenitors of type Ia 
supernovae (see, e.g., Chen et al. 2011 for a model of tidally enhanced stellar wind as a possible
pathway between a symbiotic system and a type Ia supernova), prompted by the likely increase in the WD mass after each 
outburst, further motivates the present study.

The manuscript is organized as follows: The main observational features of T CrB are 
summarized in Sect. 2. The input physics and initial conditions adopted in this work are 
presented in Sect. 3. A detailed account of the results obtained in this work is given in Sects. 4 and 5. 
Finally, a summary of the most relevant conclusions drawn from this research is provided in Sect. 6.

\section{Observational properties of T CrB}
\label{tcrb_observables}

According to Duerbeck (2009), T CrB was the first nova to be studied spectroscopically (Huggins \& Miller 1866). 
Sanford (1949) was the first to detect radial velocity variations caused by the orbital motion in T CrB, estimating an orbital period 
of 230.5 days\footnote{See Hric et al. (1997) and Schaefer (2009) for more recent determinations of the orbital period in T CrB, 
around $P_{orb} \sim 227$ days.}. Subsequently, Kraft (1958) provided precise radial velocity measurements for
both components of the binary system, enabling the determination of the orbital parameters and masses, sparking debate about the nature 
of the hot component (initially proposed to be a main-sequence star,
due to its inferred mass exceeding the Chandrasekhar limit; see, e.g., Kenyon \& Garcia 1986, Webbink et al. 1987, and references therein). 
This paradigm shifted with the analysis of satellite observations from the International Ultraviolet Explorer (IUE) by Selvelli, Cassatella \& Gilmozzi (1992), 
which suggested that the 
properties of the outburst could be better understood on the basis of a TNR on the surface of a massive WD\footnote{See Gilmozzi et al. (1991) for an alternative explanation involving a triple stellar system.}. Hric et al. (1998) 
put an end to the debate, deriving a mass of $M_{gM} = 1.38 \pm 0.2$ M$_\odot$ for the M3 giant companion
overfilling its Roche lobe, and establishing a value of $M_{\rm WD} = 1.2 \pm 0.2$ M$_\odot$ for the WD component, 
based on an assumed system inclination of $i = 68^\circ$. 
Similar values have been reported by other authors (e.g., $M_{gM} = 1.14$ M$_\odot$ and $M_{\rm WD} = 1.34$ M$_\odot$ for $i = 65^\circ$ in  
Zamanov et al. 2003; and $M_{gM} = 1.12 \pm 0.23$ M$_\odot$ and $M_{\rm WD} = 1.37 \pm 0.13$ M$_\odot$ for $i = 67^\circ$ in 
Stanishev et al. 2004; Hachisu \& Kato 1999, however, reported somewhat different values, 
$M_{gM} \sim 0.7$ M$_\odot$ and $M_{\rm WD} = 1.35$ M$_\odot$ for $i \sim 70^\circ$, assuming a tilted disk with an inclination 
of $i_{disk} \sim 35^\circ$). 
 Other estimates of the WD mass in T CrB have been reported by Hachisu \& Kato (2001) [$M_{\rm WD} = 1.37$ M$_\odot$] 
and Shara et al. (2018) [$M_{\rm WD} = 1.32$ M$_\odot$].
Regarding the WD luminosity, Zamanov et al. (2023) provide estimates for the hot component in T CrB, 
assuming a distance to the source of 914 pc\footnote{See Schaefer (2009) for a distance determination to T CrB of $800 \pm 140$ pc.}: 
during the superactive state (2016-2022), the WD was found to have an optical luminosity of 40 - 110 L$_\odot$, 
while in subsequent months, the luminosity decreased to 20 - 25 L$_\odot$ in April-May 2023 and further to 8 - 9 L$_\odot$ by August 2023.

Selvelli et al. (1992) estimated a mass-accretion rate of $\dot M_{\rm acc} = 2.32 \times 10^{-8}$ M$_\odot$ yr$^{-1}$ for T CrB during quiescence, 
consistent with the high values required to trigger a TNR on a massive WD. More recently,  Zamanov et al. (2023), inferred a broader range 
for the mass-accretion rate, $\dot M_{\rm acc} = (0.27 - 4.05) \times 10^{-8}$ M$_\odot$ yr$^{-1}$, by linking the WD optical luminosity to the 
mass-accretion rate. It is also worth noting that several studies have suggested a variation of the mass-accretion rate throughout 
the nova cycle (see, e.g., Anupama \& Prabhu 1991, Dobrotka, Hric \& Petr\'\i k 2004, Linford et al. 2019, Luna et al. 2019).

\begin{table}
\caption{Solar abundances from $^1$H to $^{48}$Ti and isotopic ratios (Lodders et al. 2009).}
\label{table1}
\centering   
    \begin{tabular}{l c}
    \hline\hline\
    X($^1$H)                              & 7.112(-1)     \\
    X($^2$H)                              & 2.760(-5)      \\
    X($^3$He)                             & 8.464(-5)     \\
    X($^4$He)                             & 2.737(-1)     \\
    X($^6$Li)                             & 6.883(-10)    \\
    X($^7$Li)                             & 9.826(-9)     \\
    X($^9$Be)                             & 1.503(-10)      \\
    X($^{10}$B)                           & 1.009(-9)  \\
    X($^{11}$B)                           & 4.529(-9)  \\
    X($^{12}$C)                           & 2.325(-3)  \\
    X($^{13}$C)                           & 2.831(-5)  \\
    X($^{14}$N)                           & 8.088(-4)  \\
    X($^{15}$N)                           & 3.180(-6)  \\
    X($^{16}$O)                           & 6.842(-3)  \\
    X($^{17}$O)                           & 2.733(-6)  \\
    X($^{18}$O)                           & 1.545(-5)  \\
    X($^{19}$F)                           & 4.162(-7) \\
    X($^{20}$Ne)                          & 1.667(-3) \\
    X($^{21}$Ne)                          & 4.193(-6) \\
    X($^{22}$Ne)                          & 1.348(-4) \\
    X($^{23}$Na)                          & 3.614(-5) \\
    X($^{24}$Mg)                          & 5.293(-4) \\
    X($^{25}$Mg)                          & 7.012(-5) \\
    X($^{26}$Mg)                          & 8.000(-5) \\
    X($^{27}$Al)                          & 6.219(-5) \\
    X($^{28}$Si)                          & 7.028(-4) \\
    X($^{29}$Si)                          & 3.695(-5) \\
    X($^{30}$Si)                          & 2.524(-5) \\
    X($^{31}$P)                           & 7.005(-6) \\
    X($^{32}$S)                           & 3.487(-4) \\
    X($^{33}$S)                           & 2.839(-6) \\
    X($^{34}$S)                           & 1.647(-5) \\
    X($^{35}$Cl)                          & 3.735(-6) \\
    X($^{36}$S)                           & 7.056(-8) \\
    X($^{36}$Ar)                          & 7.683(-5) \\
    X($^{37}$Cl)                          & 1.259(-6) \\
    X($^{38}$Ar)                          & 1.479(-5) \\
    X($^{39}$K)                           & 3.716(-6) \\
    X($^{40}$Ar)                          & 2.395(-8) \\
    X($^{40}$Ca)                          & 6.370(-5) \\
    X($^{41}$K)                           & 2.824(-7) \\
    X($^{42}$Ca)                          & 4.470(-7) \\
    X($^{43}$Ca)                          & 9.598(-8) \\
    X($^{44}$Ca)                          & 1.509(-6) \\
    X($^{45}$Sc)                          & 4.214(-8) \\
    X($^{46}$Ti)                          & 2.554(-7) \\
    X($^{47}$Ti)                          & 2.354(-7) \\
    X($^{48}$Ti)                          & 2.378(-6) \\
    \hline
    $^{12}$C/$^{13}$C                     & 82.1 \\
    $^{14}$N/$^{15}$N                     & 254 \\
    $^{16}$O/$^{17}$O                     & 2503 \\
    $^{16}$O/[$^{18}$O + $^{18}$F]        & 443 \\
    $^{28}$Si/$^{29}$Si                   & 19.0  \\
    $^{28}$Si/$^{30}$Si                   & 27.8 \\
    \hline
    \end{tabular}
\end{table}

With regard to the metallicity of the material accreted onto the WD, most studies indicate near-solar values, with some peculiar elemental features. 
Shahbaz et al. (1999) identified Li in the spectra of the giant star in T CrB, reporting an abundance of A(Li) = 12 + Log [N(Li)/N(H)] $\sim 0.6$. 
This value is a  factor of 4 below solar but significantly higher than those typically found in    stars of the same spectral type, 
where no recognizable Li lines are observed. This anomaly has been attributed to either intrinsic activity in the giant star or
contamination from the nova explosion. Similarly, Wallerstein et al. (2008) reported a Li abundance close to solar, 
with A(Li) $\sim 0.8$\footnote{A value of A(Li) $\sim 1.2$ was also reported for the RN RS Oph.}, along with an overall metallicity 
consistent with solar values. More recently, Woodward et al. (2020) reported updated lithium (Li) abundance measurements for the red giant in T CrB. 
Using a non-irradiated (classical) 1D model atmosphere, they determined A(Li) = $1.3 \pm 0.1$. When accounting for irradiation from the WD 
and/or the accretion disk, they found a higher value of A(Li) = $2.4 \pm 0.1$. Their best-fit results were also consistent with a solar metallicity 
atmosphere.
Pavlenko et al. (2020) also conducted abundance determinations for C, O, and Si in the photosphere of the red giant component of T CrB, 
complementing earlier lower-resolution spectroscopic studies by Evans et al. (2019). While the inferred elemental abundances of C, O, and Si, 
as well as some isotopic ratios, such as $^{16}$O/$^{17}$O, $^{28}$Si/$^{29}$Si, and $^{28}$Si/$^{30}$Si, align with the expected composition 
of a red giant star after the first dredge-up, notable anomalies were observed in $^{12}$C/$^{13}$C and, more strikingly, $^{16}$O/$^{18}$O. 
These deviations are inconsistent with the normal composition of a red giant, even when accounting for possible contamination from nova ejecta\footnote{See
Figueira et al. (2025) for recent 3D simulations of the impact of the ejecta on the secondary star in the RN system U Sco.}.

\section{Model and input physics}

\begin{table}
   \caption{Models of T CrB that yield $\tau_{\rm rec} = 80 \pm 0.5$ yr.}
    \label{table2}
    \centering   
    \begin{tabular}{l c c c c}
    \hline\hline\
    Model& M$_{\rm WD}$(M$_\odot$)& L$_{\rm WD}$(L$_\odot$)& Z$_{\rm acc}$(Z$_\odot$)& $\dot {\rm M}_{\rm acc}$ (M$_\odot$ yr$^{-1}$)\\
    130A & 1.30                   & 1.0                    & 1.0                 & $6.10 \times 10^{-8}$ \\
    130B & 1.30                   & 0.1                    & 1.0                 & $7.16 \times 10^{-8}$ \\
    130C & 1.30                   & 0.01                   & 1.0                 & $10.5 \times 10^{-8}$ \\
    135A & 1.35                   & 1.0                    & 1.0                 & $1.68 \times 10^{-8}$ \\
    135B & 1.35                   & 0.1                    & 1.0                 & $2.07 \times 10^{-8}$ \\
    135C & 1.35                   & 0.01                   & 1.0                 & $3.85 \times 10^{-8}$ \\
    138A & 1.38                   & 1.0                    & 1.0                 & $0.76 \times 10^{-8}$ \\
    138B & 1.38                   & 0.1                    & 1.0                 & $1.07 \times 10^{-8}$ \\
    138C & 1.38                   & 0.01                   & 1.0                 & $1.67 \times 10^{-8}$ \\
    138D & 1.38                   & 0.1                    & 0.1                 & $2.45 \times 10^{-8}$ \\
    138E & 1.38                   & 0.1                    & 10.0                & $0.36 \times 10^{-8}$ \\
    \hline
    \end{tabular}
\end{table}

The RN simulations reported in this study were performed with the hydrodynamic, Lagrangian, finite-difference, time-implicit code {\tt SHIVA} 
(see Jos\'e \& Hernanz 1998, Jos\'e 2016 for further details). 
{\tt SHIVA} relies on the standard set of differential equations of stellar evolution (i.e., mass, energy, and momentum conservation, energy transport) 
and has been extensively used in the study of classical and RN outbursts, type I X-ray bursts, and (sub-Chandrasekhar) supernova explosions, 
for over 25 years. 
The code uses a general equation of state encompassing 
contributions from the degenerate electron gas, the multicomponent ion plasma, and radiation (Blinnikov et al. 1996). 
Coulomb corrections to the electron pressure are included, and OPAL Rosseland mean opacities have been adopted (Iglesias \&  Rogers 1996).
Nuclear energy generation is obtained via a reaction network
containing 120 species (ranging from $^1$H to $^{48}$Ti), linked through 630 nuclear processes, with updated rates from the STARLIB database 
(Sallaska et al. 2013; Iliadis, priv. comm.). Screening factors and neutrino energy losses have also been taken into account.
In most simulations reported in this work, the plasma transferred from the secondary star is assumed to have solar composition (Lodders et al. 2009; 
see our Table 1) 
and is transferred at a constant rate (see, however, Sect. 4 for simulations with different metallicities).
{\tt SHIVA} also incorporates a time-dependent formalism for convective transport that sets in when the characteristic convective timescale becomes
larger than the integration time step. Partial mixing between adjacent convective shells is modeled using a diffusion equation (Prialnik et al. 1979). 
No additional semi-convection or thermohaline mixing has been included.

The large mean value of the WD mass in RNe reported in Shara et al. (2018), 1.31 M$_\odot$, justifies our choice of very large values for M$_{\rm WD}$.  
Indeed, three different values for the mass of the white dwarf hosting the thermonuclear explosion have been explored: 1.30 M$_\odot$, 
1.35 M$_\odot$, and 1.38 M$_\odot$  (see, however, Sect. 5 for 
some additional models computed with M$_{\rm WD} = 1.2$ M$_\odot$,
to account for the existence of RNe, such as T Pyx, IM Nor, and CI Aql, whose WD masses have been reported in the range 1.21 -  1.23 M$_\odot$).

For each mass, three models have been computed, corresponding to three different  
initial luminosity values of the WD: 0.01 L$_\odot$, 0.1 L$_\odot$ and L$_\odot$. A series of tests have been conducted
for each of these nine combinations of parameters, aimed at determining the mass-accretion rate at which a nova explosion 
occurs after 80 yr, the recurrence period inferred for T CrB. The corresponding models were flagged and further evolved through the 
explosion, expansion and ejection stages. For completeness, two additional models 
have been considered: while the matter transfer from the secondary star in the nine models discussed above  was assumed to be solar, 
two different cases, with 0.1 Z$_\odot$ and 10 Z$_\odot$, were computed to examine the influence of the metallicity of the accreted material on
the properties of the outburst.  It is worth noting that the secondary stars in RNe are often more evolved than those in classical novae. 
Some evolved stars display significantly enhanced surface metallicities, reaching super-solar levels. 
Observations have reported metallicities up to 2 - 3 times the solar value in some red giants (see also Cinquegrana \& Karakas 2022
for a study of asymptotic giant branch models with metallicities ranging from 3 Z$_\odot$ to 7 Z$_\odot$). In contrast, binary systems 
originating in the Galactic halo or other metal-poor environments, such as the Magellanic clouds and dwarf galaxies, are likely to 
contain metal-deficient stars. Within this context, the adopted metallicity range (0.1 - 10) Z$_\odot$ must be regarded as exploratory, 
representing a variation of the solar metallicity by a factor of 10 up and down, in an attempt to quantify the role played by the metallicity 
of the companion star on the properties of the outbursts.
All in all, 11 new hydrodynamic simulations have been performed for this study (see Table 2).

\begin{table}
   \caption{Properties of T CrB models with M$_{\rm WD} = 1.30$ M$_\odot$.}
    \label{table3}
    \centering   
    \begin{tabular}{l c c c}
    \hline\hline\
    Model                                 & 130A                 & 130B                 & 130C      \\                  
    M$_{\rm WD}$(M$_\odot$)               & 1.30                 & 1.30                 & 1.30      \\ 
    L$_{\rm WD}$(L$_\odot$)               & 1.0                  & 0.1                  & 0.01      \\ 
    Z$_{\rm acc}$                         & 0.014 (Z$_\odot$)    & 0.014 (Z$_\odot$)    & 0.014 (Z$_\odot$)   \\
    $\dot {\rm M}_{\rm acc}$  
                (M$_\odot$ yr$^{-1}$)     &$6.10 \times 10^{-8}$ &$7.16 \times 10^{-8}$ &$1.05 \times 10^{-7}$ \\
    \hline
    $\Delta$M$_{\rm acc}$($10^{-6}$ M$_\odot$)& 5.40                 & 6.24                 & 8.66  \\
    $\rho_{max}$($10^3$ g cm$^{-3}$)      & 2.16                 & 2.53                 & 3.93  \\
    T$_{max}$($10^8$ K)                   & 2.01                 & 2.08                 & 2.22  \\
    P$_{max}$($10^{19}$ dyn cm$^{-2}$)    & 1.17                 & 1.35                 & 1.88  \\
    K($10^{44}$ ergs)                     & 2.55                 & 2.85                 & 3.91   \\
    v(km s$^{-1}$)                        & 2502                 & 2459                 & 2301   \\
    $\Delta$M$_{eje}$($10^{-6}$ M$_\odot$)& 3.53                 & 4.07                 & 6.29      \\
    Z$_{eje}$                             & 0.014                & 0.014                & 0.014     \\
    \hline
    \end{tabular}
\end{table}

\section{Results}

\subsection{Effects of the WD mass and luminosity, the metallicity, and the accretion rate} 
In a nova explosion, mass ejection requires a critical pressure at the envelope's base, P$_{\rm cri}$, which is approximately given as  
\begin{equation}
P_{\rm cri} = \frac{G \, M_{\rm WD}}{4 \pi R_{\rm WD}^4} \Delta M_{\rm acc} ~,
\end{equation}
where M$_{\rm WD}$ and R$_{\rm WD}$ are the mass and radius 
of the WD hosting the explosion, and $\Delta$M$_{\rm acc}$ is the mass of the 
accreted envelope (Shara 1981, Fujimoto 1982).  Pressures within the range
P$_{\rm cri} \sim 10^{19} - 10^{20}$ dyn cm$^{-2}$ must be attained,
the exact value depending upon the chemical composition (Fujimoto 1982, MacDonald 1983; see also Wolf et al. 2013 and Kato et al. 2014 for more recent work on ignition masses). Higher ignition pressures lead to more violent outbursts, characterized by higher peak temperatures and a somewhat larger extension of the nuclear activity toward heavier species.

\begin{table}
   \caption{Properties of T CrB models with M$_{\rm WD} = 1.35$ M$_\odot$.}
    \label{table4}
    \centering   
    \begin{tabular}{l c c c}
    \hline\hline\
    Model                                 & 135A                 & 135B                 & 135C      \\                  
    M$_{\rm WD}$(M$_\odot$)               & 1.35                 & 1.35                 & 1.35      \\ 
    L$_{\rm WD}$(L$_\odot$)               & 1.0                  & 0.1                  & 0.01      \\ 
    Z$_{\rm acc}$                         & 0.014 (Z$_\odot$)    & 0.014 (Z$_\odot$)    & 0.014 (Z$_\odot$)   \\
    $\dot {\rm M}_{\rm acc}$  
                (M$_\odot$ yr$^{-1}$)     &$1.68 \times 10^{-8}$ &$2.07 \times 10^{-8}$ &$3.85 \times 10^{-8}$ \\
    \hline
    $\Delta$M$_{\rm acc}$($10^{-6}$ M$_\odot$)& 1.82                 & 2.14                 & 3.49  \\
    $\rho_{max}$($10^3$ g cm$^{-3}$)      & 3.43                 & 4.02                 & 7.01  \\
    T$_{max}$($10^8$ K)                   & 2.36                 & 2.45                 & 2.71  \\
    P$_{max}$($10^{19}$ dyn cm$^{-2}$)    & 1.90                 & 2.23                 & 3.66  \\
    K($10^{44}$ ergs)                     & 1.73                 & 1.99                 & 2.98   \\
    v(km s$^{-1}$)                        & 3558                 & 3531                 & 3361   \\
    $\Delta$M$_{eje}$($10^{-6}$ M$_\odot$)& 1.19                 & 1.40                 & 2.29      \\
    Z$_{eje}$                             & 0.015                & 0.015                & 0.015     \\
    \hline
    \end{tabular}
\end{table}

\begin{table*}[tb]
   \caption{Properties of T CrB models with M$_{\rm WD} = 1.38$ M$_\odot$.}
    \label{table5}
    \centering   
    \begin{tabular}{l c c c c c}
    \hline\hline\
    Model                                 & 138A                 & 138B                 & 138C                 & 138D                 & 138E \\                  
    M$_{\rm WD}$(M$_\odot$)               & 1.38                 & 1.38                 & 1.38                 & 1.38                 & 1.38 \\ 
    L$_{\rm WD}$(L$_\odot$)               & 1.0                  & 0.1                  & 0.01                 & 0.1                  & 0.1  \\ 
    Z$_{\rm acc}$                         & 0.014 (Z$_\odot$)    & 0.014 (Z$_\odot$)    & 0.014 (Z$_\odot$)    & 0.0014 (0.1 Z$_\odot$)  & 0.14 (10 Z$_\odot$) \\
    $\dot {\rm M}_{\rm acc}$  
                (M$_\odot$ yr$^{-1}$)     &$7.57 \times 10^{-9}$ &$1.07 \times 10^{-8}$ &$1.67 \times 10^{-8}$ &$2.45 \times 10^{-8}$ &$3.55 \times 10^{-9}$ \\
    \hline
    $\Delta$M$_{\rm acc}$($10^{-6}$ M$_\odot$)&  0.808               & 1.06                 & 1.48                 & 2.62                 & 0.525 \\
    $\rho_{max}$($10^3$ g cm$^{-3}$)      &  4.19                & 5.38                 & 8.12                 & 11.4                 & 3.69  \\
    T$_{max}$($10^8$ K)                   &  2.54                & 2.71                 & 2.92                 & 3.25                 & 2.33  \\
    P$_{max}$($10^{19}$ dyn cm$^{-2}$)    &  2.39                & 3.12                 & 4.40                 & 7.72                 & 1.56  \\
    K($10^{44}$ ergs)                     &  1.17                & 1.44                 & 1.93                 & 2.80                 & 1.03  \\
    v(km s$^{-1}$)                        &  4649                & 4553                 & 4093                 & 3998                 & 4828  \\
    $\Delta$M$_{eje}$($10^{-6}$ M$_\odot$)& 0.474                & 0.619                & 0.972                & 1.54                 &  0.381\\
    Z$_{eje}$                             & 0.014                & 0.015                & 0.015                & 0.0015               &  0.14 \\
    \hline
    \end{tabular}
\end{table*}

As shown in Table 3, 1.30 M$_\odot$ WDs with an initial luminosity ranging from 0.01 to 1 L$_\odot$,
require mass-accretion rates of around $10^{-7}$ M$_\odot$ yr$^{-1}$
to drive a nova explosion after $\sim 80$ yr of accretion 
of solar-composition matter. Specifically, a mass-accretion rate of $1.05 \times 10^{-7}$ M$_\odot$ yr$^{-1}$ is required for L$_{\rm WD} = 0.01$ L$_\odot$, while $6.1 \times 10^{-8}$ M$_\odot$ yr$^{-1}$ is needed for L$_{\rm WD} = 1$ L$_\odot$, showing that the lower the initial WD luminosity, the higher the mass-accretion rate required to power such an explosion in 80 yr. 

It is well-known that hotter (i.e., more luminous) accreting WDs attain the critical conditions to initiate a TNR earlier, 
thus reducing the duration of the accretion phase and the overall amount of mass accreted. A similar outcome is obtained in models 
characterized by higher mass-accretion rates. Accordingly, to produce an outburst after 80 yr, models with higher initial WD luminosities require lower mass-accretion rates to compensate for the reduced amount of accreted mass,  in agreement with the 
results summarized in Table 3.

Increasing the WD mass to 1.35 M$_\odot$ (Table 4) results in a similar pattern\footnote{See also Table 5
for 1.38 M$_\odot$ WDs.}, albeit with lower 
mass-accretion rates (a decrease by a factor of 3, compared to models computed with 1.30 M$_\odot$).
This can be understood as follows: replacing the accreted mass by the product of the 
mass-accretion rate, $\dot M_{\rm acc}$, and the duration of the accretion phase, $t_{\rm acc}$, Eq. (1) becomes 
\begin{equation}
P_{\rm cri} \propto \frac{M_{\rm WD} \dot M_{\rm acc} t_{\rm acc}}{R_{\rm WD}^4} 
.\end{equation}
Taking into account that in all models reported in this work $t_{\rm acc} = \tau_{\rm rec} = 80$ yr (i.e., $t_{\rm acc} = const.$), we get
\begin{equation}
P_{\rm cri} \propto \frac{M_{\rm WD} \dot M}{R_{\rm WD}^4} 
.\end{equation}
And recalling that more massive WDs are also more compact (i.e., they have a smaller radius), in order to achieve the same $P_{\rm cri}$ at the base of the envelope, 
a more massive WD requires a lower $\dot M$, as shown in Tables 3 and 4.

Finally, an increase in the metallicity of the accreted material translates into an overall reduction in the
mass-accretion rate required to achieve $\tau_{\rm rec} \sim 80$ yr (see models 138D, 138B, and 138E in Table 5).
A higher metallicity favors an earlier occurrence of the TNR,  
due to the larger number of nuclear reactions, in particular $^{12}$C(p, $\gamma$) (see Jos\'e, Halabi \& El Eid 2016), 
thereby shortening the overall duration of the accretion phase and decreasing the amount of mass accreted in the envelope. 
Accordingly,  for a fixed $t_{\rm acc} = \tau_{\rm rec} = 80$ yr, models with higher metallicity require lower mass-accretion 
rates to compensate for the smaller accreted mass.

Tables 3 to 5 reveal a consistent trend: given that the accretion phase duration is fixed to 80 yr in all models, it turns out that higher accreted masses, $\Delta M_{\rm acc} = \dot M_{\rm acc} t_{\rm acc}$, are obtained in models evolved with lower $L_{\rm WD}$ or $Z$ (i.e., models requiring higher mass-accretion rates to trigger a nova explosion after $t_{\rm acc}$ = 80 yr), which in turn translates into larger ignition densities and pressures in the envelope. Since pressure at the base of the envelope determines the strength of the explosion, it follows that models with lower $L_{\rm WD}$ or $Z$ (including those with lower $M_{\rm WD}$) result in more violent outbursts, characterized by higher 
peak temperatures ($T_{max}$) and larger ejected masses ($\Delta M_{\rm eje}$) with greater kinetic energies ($K$).

\subsection{Nucleosynthesis} 
Nucleosynthesis provides valuable observables that can help identify the model that better matches the explosion of T CrB.

\subsubsection{Isotopic abundances}
Spectroscopic observations of the nova ejecta often provide elemental abundances only. However, a number of isotopic ratios have been accessible through high-resolution, near-IR spectroscopy (eg., $^{12}$C/$^{13}$C through the first
overtone of CO in absorption; see Evans et al. 2019, and Pavlenko et al. 2020 for details\footnote{See also Evans et al. (2025) for the first near-
IR spectroscopy of an extragalactic nova, the RN LMCN 1968-12a.}). Here, we review the most important results in terms of mean (mass-averaged) 
isotopic abundances in the ejecta of the different models (see Tables 6 to 10; tables with complete yields can be found in the appendix).

\begin{table}
   \caption{Isotopic abundances$^a$ of the CNOF-group nuclei in the ejecta of the main T CrB models.}
    \label{table6}
    \centering   
    \begin{tabular}{l c c c}
    \hline\hline\
    L$_{\rm WD}$(L$_\odot$)       & 1                    & 0.1                  & 0.01     \\
    \hline
    M$_{\rm WD}$ = 1.30 M$_\odot$ & 130A                 & 130B                 & 130C      \\ 
    X($^{12}$C)                   & 1.5(-3)              & 1.6(-3)              & 1.8(-3)   \\
    X($^{13}$C)                   & 2.5(-3)              & 2.6(-3)              & 2.4(-3)   \\
    X($^{14}$N)                   & 4.5(-3)              & 4.0(-3)              & 3.5(-3)   \\
    X($^{15}$N)                   & 7.1(-4)              & 8.9(-4)              & 1.5(-3)   \\
    X($^{16}$O)                   & 2.2(-5)              & 1.9(-5)              & 1.4(-5)   \\
    X($^{17}$O)                   & 3.8(-7)              & 4.3(-7)              & 5.9(-7)   \\
    X($^{18}$O)$^b$               & -                    & -                    & -         \\
    X($^{18}$F)$^b$               & 2.9(-9)              & 2.3(-9)              & 2.0(-9)   \\
    X($^{19}$F)                   & -                    & -                    & -         \\
    \hline
    M$_{\rm WD}$ = 1.35 M$_\odot$ & 135A                 & 135B                 & 135C      \\ 
    X($^{12}$C)                   & 1.6(-3)              & 1.7(-3)              & 1.7(-3)   \\
    X($^{13}$C)                   & 2.3(-3)              & 2.0(-3)              & 1.4(-3)   \\
    X($^{14}$N)                   & 3.2(-3)              & 3.2(-3)              & 3.4(-3)   \\
    X($^{15}$N)                   & 2.2(-3)              & 2.4(-3)              & 2.8(-3)   \\
    X($^{16}$O)                   & 1.3(-5)              & 1.1(-5)              & 6.0(-6)   \\
    X($^{17}$O)                   & 7.6(-7)              & 8.4(-7)              & 9.7(-7)   \\
    X($^{18}$O)$^a$               & -                    & -                    & -  \\
    X($^{18}$F)$^a$               & 2.4(-9)              & 2.3(-9)              & 2.5(-9)   \\
    X($^{19}$F)                   & -                    & -                    & -         \\
    \hline
    M$_{\rm WD}$ = 1.38 M$_\odot$ & 138A                 & 138B                 & 138C      \\ 
    X($^{12}$C)                   & 1.6(-3)              & 1.5(-3)              & 1.6(-3)   \\
    X($^{13}$C)                   & 1.9(-3)              & 1.5(-3)              & 1.2(-3)   \\
    X($^{14}$N)                   & 3.2(-3)              & 3.3(-3)              & 3.4(-3)   \\
    X($^{15}$N)                   & 2.8(-3)              & 3.1(-3)              & 3.3(-3)   \\
    X($^{16}$O)                   & 9.5(-6)              & 6.6(-6)              & 4.4(-6)    \\
    X($^{17}$O)                   & 8.7(-7)              & 9.7(-7)              & 1.0(-6)    \\
    X($^{18}$O)$^a$               & -                    & -                    & -          \\
    X($^{18}$F)$^a$               & 2.7(-9)              & 2.9(-9)              & 3.6(-9)   \\
    X($^{19}$F)                   & -                    & -                    & -          \\
    \hline
    \end{tabular}
\vspace{0.1 cm}
\begin{list}{}{} 
\item[$^{\mathrm{a}}$] Mass fractions listed for $> 10^{-10}$.
\item[$^{\mathrm{b}}$] Mass fractions correspond to t = 1 hr after T$_{max}$.
\end{list}
\end{table}

\begin{table}
   \caption{Isotopic abundances$^a$ of Ne and Na nuclei in the ejecta of the main T CrB models.}
    \label{table7}
    \centering   
    \begin{tabular}{l c c c}
    \hline\hline\
    L$_{\rm WD}$(L$_\odot$)       & 1                    & 0.1                  & 0.01     \\
    \hline
    M$_{\rm WD}$ = 1.30 M$_\odot$ & 130A                 & 130B                 & 130C      \\ 
    X($^{20}$Ne)                  & 1.6(-3)              & 1.5(-3)              & 1.2(-3)   \\
    X($^{21}$Ne)                  & 1.6(-7)              & 1.7(-7)              & 1.7(-7)   \\
    X($^{22}$Ne)                  & 6.4(-7)              & 8.8(-8)              & 1.7(-10)  \\
    X($^{22}$Na)                  & 1.9(-6)              & 1.7(-6)              & 1.3(-6)   \\
    X($^{23}$Na)                  & 1.4(-6)              & 1.3(-6)              & 1.3(-6)   \\
    \hline
    M$_{\rm WD}$ = 1.35 M$_\odot$ & 135A                 & 135B                 & 135C      \\ 
    X($^{20}$Ne)                  & 6.5(-4)              & 4.4(-4)              & 2.2(-4)   \\
    X($^{21}$Ne)                  & 1.2(-7)              & 8.6(-8)              & 4.5(-8)   \\
    X($^{22}$Ne)                  & -                    & -                    & -         \\
    X($^{22}$Na)                  & 6.8(-7)              & 4.9(-7)              & 3.6(-7)   \\
    X($^{23}$Na)                  & 7.3(-7)              & 6.1(-7)              & 6.6(-7)   \\
    \hline
    M$_{\rm WD}$ = 1.38 M$_\odot$ & 138A                 & 138B                 & 138C      \\ 
    X($^{20}$Ne)                  & 5.1(-4)              & 2.4(-4)              & 1.3(-4)   \\
    X($^{21}$Ne)                  & 1.1(-7)              & 5.5(-8)              & 3.0(-8)   \\
    X($^{22}$Ne)                  & -                    & -                    & -         \\
    X($^{22}$Na)                  & 5.8(-7)              & 3.7(-7)              & 3.1(-7)   \\
    X($^{23}$Na)                  & 7.8(-7)              & 6.2(-7)              & 6.8(-7)   \\
    \hline
    \end{tabular}
\vspace{0.1 cm}
\begin{list}{}{} 
\item[$^{\mathrm{a}}$] Mass fractions listed for $> 10^{-10}$.
\end{list}
\end{table}

\begin{table}
   \caption{Isotopic abundances$^a$ of Mg, Al, and Si nuclei in the ejecta of the main T CrB models.}
    \label{table8}
    \centering   
    \begin{tabular}{l c c c}
    \hline\hline\
    L$_{\rm WD}$(L$_\odot$)       & 1                    & 0.1                  & 0.01     \\
    \hline
    M$_{\rm WD}$ = 1.30 M$_\odot$ & 130A                 & 130B                 & 130C      \\ 
    X($^{24}$Mg)                  & 3.3(-8)              & 3.1(-8)              & 2.7(-8)   \\
    X($^{25}$Mg)                  & 6.4(-6)              & 5.5(-6)              & 3.6(-6)   \\
    X($^{26}$Mg)                  & 2.6(-7)              & 2.3(-7)              & 1.5(-7)   \\
    X($^{26g}$Al)                 & 1.1(-6)              & 1.1(-6)              & 8.9(-7)   \\
    X($^{27}$Al)                  & 6.1(-6)              & 5.9(-6)              & 4.7(-6)   \\
    X($^{28}$Si)                  & 1.6(-3)              & 1.5(-3)              & 9.5(-4)   \\
    X($^{29}$Si)                  & 1.0(-5)              & 9.8(-6)              & 6.9(-6)   \\
    X($^{30}$Si)                  & 3.0(-4)              & 4.4(-4)              & 6.6(-4)   \\
    \hline
    M$_{\rm WD}$ = 1.35 M$_\odot$ & 135A                 & 135B                 & 135C      \\ 
    X($^{24}$Mg)                  & 1.6(-8)              & 1.2(-8)              & 1.1(-8)   \\
    X($^{25}$Mg)                  & 1.6(-6)              & 1.0(-6)              & 7.1(-7)   \\
    X($^{26}$Mg)                  & 6.8(-8)              & 4.4(-8)              & 3.0(-8)   \\
    X($^{26g}$Al)                 & 5.1(-7)              & 3.5(-7)              & 2.0(-7)   \\
    X($^{27}$Al)                  & 2.7(-6)              & 1.8(-6)              & 1.1(-6)   \\
    X($^{28}$Si)                  & 3.2(-4)              & 2.0(-4)              & 9.5(-5)   \\
    X($^{29}$Si)                  & 2.5(-6)              & 1.7(-6)              & 1.2(-6)   \\
    X($^{30}$Si)                  & 3.1(-4)              & 1.8(-4)              & 7.7(-5)   \\
    \hline
    M$_{\rm WD}$ = 1.38 M$_\odot$ & 138A                 & 138B                 & 138C      \\ 
    X($^{24}$Mg)                  & 1.5(-8)              & 1.1(-8)              & 1.0(-8)    \\
    X($^{25}$Mg)                  & 1.1(-6)              & 6.4(-7)              & 6.0(-7)   \\
    X($^{26}$Mg)                  & 5.0(-8)              & 2.8(-8)              & 2.6(-8)   \\
    X($^{26g}$Al)                 & 4.1(-7)              & 2.1(-7)              & 1.5(-7)   \\
    X($^{27}$Al)                  & 2.2(-6)              & 1.2(-6)              & 8.3(-7)   \\
    X($^{28}$Si)                  & 2.9(-4)              & 1.1(-4)              & 4.5(-5)   \\
    X($^{29}$Si)                  & 2.8(-6)              & 1.3(-6)              & 9.4(-7)   \\
    X($^{30}$Si)                  & 3.8(-4)              & 1.0(-4)              & 3.5(-5)   \\
    \hline
    \end{tabular}
\vspace{0.1 cm}
\begin{list}{}{} 
\item[$^{\mathrm{a}}$] Mass fractions listed for $> 10^{-10}$.
\end{list}
\end{table}

\begin{table}
   \caption{Isotopic abundances$^a$ of P and S nuclei in the ejecta of the main T CrB models. }
    \label{table9}
    \centering   
    \begin{tabular}{l c c c}
    \hline\hline\
    L$_{\rm WD}$(L$_\odot$)       & 1                    & 0.1                  & 0.01     \\
    \hline
    M$_{\rm WD}$ = 1.30 M$_\odot$ & 130A                 & 130B                 & 130C      \\ 
    X($^{31}$P)                   & 3.0(-5)              & 5.2(-5)              & 8.3(-5)   \\
    X($^{32}$S)                   & 3.9(-4)              & 4.8(-4)              & 1.4(-3)   \\
    X($^{33}$S)                   & 1.6(-6)              & 1.2(-6)              & 1.1(-6)   \\
    X($^{34}$S)                   & 8.0(-6)              & 5.3(-6)              & 1.3(-6)   \\
    X($^{36}$S)                   & 2.2(-8)              & 1.1(-8)              & 5.0(-10)  \\
    \hline
    M$_{\rm WD}$ = 1.35 M$_\odot$ & 135A                 & 135B                 & 135C      \\ 
    X($^{31}$P)                   & 3.6(-5)              & 2.3(-5)              & 1.6(-5)   \\
    X($^{32}$S)                   & 3.3(-3)              & 3.9(-3)              & 3.3(-3)   \\
    X($^{33}$S)                   & 4.8(-6)              & 6.3(-6)              & 8.2(-6)   \\
    X($^{34}$S)                   & 4.0(-6)              & 5.2(-6)              & 5.8(-6)   \\
    X($^{36}$S)                   & -                    & -                    & -         \\
    \hline
    M$_{\rm WD}$ = 1.38 M$_\odot$ & 138A                 & 138B                 & 138C      \\ 
    X($^{31}$P)                   & 4.9(-5)              & 1.9(-5)              & 1.1(-5)   \\
    X($^{32}$S)                   & 3.0(-3)              & 3.2(-3)              & 2.4(-3)   \\
    X($^{33}$S)                   & 4.7(-6)              & 8.2(-6)              & 1.1(-5)   \\
    X($^{34}$S)                   & 3.3(-6)              & 5.5(-6)              & 6.7(-6)   \\
    X($^{36}$S)                   & -                    & -                    & -         \\
    \hline
    \end{tabular}
\vspace{0.1 cm}
\begin{list}{}{} 
\item[$^{\mathrm{a}}$] Mass fractions listed for $> 10^{-10}$.
\end{list}
\end{table}

\begin{table}
   \caption{Isotopic abundances$^a$ of Ca and Sc nuclei in the ejecta of the main T CrB models. }
    \label{table10}
    \centering   
    \begin{tabular}{l c c c}
    \hline\hline\
    L$_{\rm WD}$(L$_\odot$)       & 1                    & 0.1                  & 0.01     \\
    \hline
    M$_{\rm WD}$ = 1.30 M$_\odot$ & 130A                 & 130B                 & 130C      \\ 
    X($^{40}$Ca)                  & 6.4(-5)              & 6.4(-5)              & 6.4(-5)   \\
    X($^{42}$Ca)                  & 4.5(-7)              & 4.5(-7)              & 4.5(-7)   \\
    X($^{43}$Ca)                  & 9.7(-8)              & 9.8(-8)              & 1.0(-7)   \\
    X($^{44}$Ca)                  & 1.5(-6)              & 1.5(-6)              & 1.5(-6)   \\
    X($^{45}$Sc)                  & 4.6(-8)              & 4.8(-8)              & 6.2(-8)   \\
    \hline
    M$_{\rm WD}$ = 1.35 M$_\odot$ & 135A                 & 135B                 & 135C      \\ 
    X($^{40}$Ca)                  & 6.7(-5)              & 8.6(-5)              & 1.3(-3)   \\
    X($^{42}$Ca)                  & 4.6(-7)              & 4.5(-7)              & 5.6(-7)   \\
    X($^{43}$Ca)                  & 1.2(-7)              & 1.4(-7)              & 4.4(-7)   \\
    X($^{44}$Ca)                  & 1.5(-6)              & 1.4(-6)              & 1.2(-6)   \\
    X($^{45}$Sc)                  & 1.1(-7)              & 1.6(-7)              & 4.3(-7)   \\
    \hline
    M$_{\rm WD}$ = 1.38 M$_\odot$ & 138A                 & 138B                 & 138C      \\ 
    X($^{40}$Ca)                  & 4.3(-4)              & 1.3(-3)              & 2.8(-3)   \\
    X($^{42}$Ca)                  & 4.2(-7)              & 5.8(-7)              & 1.5(-6)   \\
    X($^{43}$Ca)                  & 1.8(-7)              & 4.9(-7)              & 3.7(-6)   \\
    X($^{44}$Ca)                  & 1.3(-6)              & 1.2(-6)              & 6.9(-6)   \\
    X($^{45}$Sc)                  & 2.5(-7)              & 4.2(-7)              & 7.6(-7)   \\
    \hline
    \end{tabular}
\vspace{0.1 cm}
\begin{list}{}{} 
\item[$^{\mathrm{a}}$] Mass fractions listed for $> 10^{-10}$.
\end{list}
\end{table}

The three models computed for $M_{\rm WD} = 1.3$ M$_\odot$ yield identical (mean) mass fractions of H and $^4$He in the ejecta. Remarkably, the H content in the envelope only experienced a minor decrease of 
8\% through the entire outburst, while X(H)+X($^4$He) remained nearly constant. Indeed, very small leakage from the CNO region is observed 
($\sum X(CNO)_{\rm ini} = 10^{-2} \rightarrow \sum X(CNO)_{\rm eje} = 
9.2 \times 10^{-3}$), indicating limited nuclear activity during the explosion. Aside from H and $^4$He, the most abundant 
isotopes in the ejecta correspond to the CNO region, including $^{14}$N, $^{12,13}$C (occasionally $^{15}$N), $^{20}$Ne, 
and in some models $^{28}$Si and $^{32}$S, all characterized by mass fractions around $\sim 10^{-3}$.
Predicted isotopic ratios\footnote{Mass fractions have been used to evaluate the corresponding isotopic ratios.} in the ejecta
for some important species include $^{12}$C/$^{13}$C = 0.60 - 0.75, $^{16}$O/$^{17}$O = 24 - 58, $^{16}$O/[$^{18}$O + $^{18}$F] = 7000 - 8300, 
$^{28}$Si/$^{29}$Si = 140 - 160, and $^{28}$Si/$^{30}$Si = 1.4 - 5.3. With regard to $^7$Li(+$^7$Be), mass fractions around $10^{-12}$ have also been obtained for these same models, revealing depletion with respect to solar values.

The $M_{\rm WD} = 1.35$ M$_\odot$ models follow a similar pattern, displaying a notable presence of $^{14,15}$N, $^{12,13}$C, and $^{32}$S in the ejecta. However, the higher peak temperature achieved in Model 135C, $T_{\rm max} = 2.71 \times 10^8$ K, resulted in an extension of the nuclear activity. Indeed, $^{40}$Ca reached a mass fraction of $1.3 \times 10^{-3}$ in the ejecta, a value comparable to those of the most abundant CNO isotopes.
With regard to isotopic ratios, these models yielded 
$^{12}$C/$^{13}$C = 0.70 - 1.2, 
$^{16}$O/$^{17}$O = 6.2 - 17, 
$^{16}$O/[$^{18}$O + $^{18}$F] = 2400 - 5400, 
$^{28}$Si/$^{29}$Si = 79 - 130, and 
$^{28}$Si/$^{30}$Si = 1.0 - 1.2, 
while $^7$Li(+$^7$Be) mass fractions reached values ranging between $10^{-12}$ - $10^{-11}$. 

Similar trends are found for the $M_{\rm WD} = 1.38$ M$_\odot$ models. The adopted mass is enough to expose the plasma to high peak temperatures 
(i.e., $2.92 \times 10^8$ K in Model 138C), which reinforced the presence of $^{40}$Ca in the ejecta, reaching $2.8 \times 10^{-3}$ in Model 138C 
(about 44 times the solar value).  With regard to isotopic ratios, the 1.38 M$_\odot$ models with solar metallicity yielded 
$^{12}$C/$^{13}$C = 0.84 - 1.3, 
$^{16}$O/$^{17}$O = 4.4 - 11, 
$^{16}$O/[$^{18}$O + $^{18}$F] = 1200 - 3500, 
$^{28}$Si/$^{29}$Si = 48 - 104, and 
$^{28}$Si/$^{30}$Si = 0.76 - 1.3, 
while $^7$Li(+$^7$Be) mass fractions reached values around $10^{-11}$.

Larger variations in chemical composition within the ejecta are observed among the additional 1.38 M$_\odot$ models evolved with different initial 
metallicities of the accreted matter: indeed, while the ejecta in Model 138D (Z = 0.1 Z$_\odot$) is characterized by significant amounts of $^{40}$Ca 
(the third, most abundant species, after H and $^4$He), $^{14,15}$N, $^{12,13}$C, and $^{44}$Ca, 
dominance shifts toward the presence of $^{14,15}$N, $^{12,13}$C, $^{20}$Ne, and $^{28}$Si in Model 138E (Z = 10 Z$_\odot$).

\subsubsection{Elemental abundances}
While some isotopic abundances can be obtained in the near IR (see Sect. 4.2.1), 
spectroscopic determinations often provide elemental (rather than isotopical) 
abundances. The main elemental abundances in the ejecta of T CrB, predicted for the
different models reported in this work, are summarized in Tables 11, 12, and 13. 

\begin{table}
   \caption{Elemental abundances for T CrB models$^a$ with M$_{\rm WD} = 1.30$ M$_\odot$.}
    \label{Table11}
    \centering   
    \begin{tabular}{l c c c}
    \hline\hline\
    Model                                 & 130A                 & 130B                 & 130C      \\                  
    M$_{\rm WD}$(M$_\odot$)               & 1.30                 & 1.30                 & 1.30      \\ 
    L$_{\rm WD}$(L$_\odot$)               & 1.0                  & 0.1                  & 0.01      \\ 
    Z$_{\rm acc}$                         & 0.014 (Z$_\odot$)    & 0.014 (Z$_\odot$)    & 0.014 (Z$_\odot$)   \\
    \hline
    Z$_{eje}$                             & 0.014                & 0.014                & 0.014     \\
    X(H)                                  & 6.5(-1)              & 6.5(-1)              & 6.5(-1)   \\
    X(He)                                 & 3.4(-1)              & 3.4(-1)              & 3.4(-1)   \\
    X(Li)                                 & -                    & -                    & -         \\
    X(Be)                                 & -                    & -                    & -          \\
    X(B)                                  & -                    & -                    & -          \\
    X(C)                                  & 3.9(-3)              & 4.2(-3)              & 4.2(-3)    \\
    X(N)                                  & 5.2(-3)              & 4.9(-3)              & 5.0(-3)    \\
    X(O)                                  & 2.3(-5)              & 2.0(-5)              & 1.5(-5)    \\
    X(F)                                  & 2.9(-9)              & 2.3(-9)              & 2.0(-9)    \\
    X(Ne)                                 & 1.6(-3)              & 1.5(-3)              & 1.2(-3)    \\
    X(Na)                                 & 3.3(-6)              & 3.0(-6)              & 2.6(-6)    \\
    X(Mg)                                 & 6.7(-6)              & 5.7(-6)              & 3.8(-6)    \\
    X(Al)                                 & 7.2(-6)              & 6.9(-6)              & 5.6(-6)    \\
    X(Si)                                 & 1.9(-3)              & 2.0(-3)              & 1.6(-3)    \\
    X(P)                                  & 3.0(-5)              & 5.2(-5)              & 8.3(-5)    \\
    X(S)                                  & 4.0(-4)              & 4.9(-4)              & 1.4(-3)    \\
    X(Cl)                                 & 6.1(-5)              & 7.7(-5)              & 8.5(-5)    \\
    X(Ar)                                 & 4.7(-5)              & 3.4(-5)              & 2.8(-5)    \\
    X(K)                                  & 4.9(-6)              & 5.5(-6)              & 9.8(-6)    \\
    X(Ca)                                 & 6.6(-5)              & 6.6(-5)              & 6.6(-5)    \\
    X(Sc)                                 & 4.6(-8)              & 4.8(-8)              & 6.2(-8)    \\
    X(Ti)                                 & 2.9(-6)              & 2.9(-6)              & 2.9(-6)    \\
    \hline
    \end{tabular}
\vspace{0.1 cm}
\begin{list}{}{} 
\item[$^{\mathrm{a}}$] Mass fractions listed for $> 10^{-10}$.
\end{list}
\end{table}

\begin{table}
   \caption{Elemental abundances for T CrB models with M$_{\rm WD} = 1.35$ M$_\odot$.}
    \label{Table12}
    \centering   
    \begin{tabular}{l c c c}
    \hline\hline\
    Model                                 & 135A                 & 135B                 & 135C      \\                  
    M$_{\rm WD}$(M$_\odot$)               & 1.35                 & 1.35                 & 1.35      \\ 
    L$_{\rm WD}$(L$_\odot$)               & 1.0                  & 0.1                  & 0.01      \\ 
    Z$_{\rm acc}$                         & 0.014 (Z$_\odot$)    & 0.014 (Z$_\odot$)    & 0.014 (Z$_\odot$)   \\
    \hline
    Z$_{eje}$                             & 0.015                & 0.015                & 0.015     \\
    X(H)                                  & 6.2(-1)              & 6.2(-1)              & 6.2(-1)          \\
    X(He)                                 & 3.6(-1)              & 3.6(-1)              & 3.6(-1)          \\
    X(Li)                                 & -                    & -                    & -    \\
    X(Be)                                 & -                    & -                    & -     \\
    X(B)                                  & -                    & -                    & -    \\
    X(C)                                  & 3.9(-3)              & 3.7(-3)              & 3.1(-3)    \\
    X(N)                                  & 5.4(-3)              & 5.6(-3)              & 6.2(-3)   \\
    X(O)                                  & 1.4(-5)              & 1.2(-5)              & 6.9(-6)    \\
    X(F)                                  & 2.4(-9)              & 2.3(-9)              & 2.5(-9)    \\
    X(Ne)                                 & 6.5(-4)              & 4.4(-4)              & 2.2(-4)    \\
    X(Na)                                 & 1.4(-6)              & 1.1(-6)              & 1.0(-6)    \\
    X(Mg)                                 & 1.7(-6)              & 1.1(-6)              & 7.5(-7)    \\
    X(Al)                                 & 3.2(-6)              & 2.2(-6)              & 1.3(-6)     \\
    X(Si)                                 & 6.3(-4)              & 3.7(-4)              & 1.7(-4)    \\
    X(P)                                  & 3.6(-5)              & 2.3(-5)              & 1.6(-5)    \\
    X(S)                                  & 3.3(-3)              & 3.9(-3)              & 3.3(-3)    \\
    X(Cl)                                 & 7.6(-5)              & 1.0(-4)              & 1.3(-4)    \\
    X(Ar)                                 & 4.2(-5)              & 6.3(-5)              & 9.5(-5)    \\
    X(K)                                  & 2.8(-5)              & 6.3(-5)              & 2.1(-4)    \\
    X(Ca)                                 & 6.9(-5)              & 8.8(-5)              & 1.3(-3)    \\
    X(Sc)                                 & 1.1(-7)              & 1.6(-7)              & 4.3(-7)    \\
    X(Ti)                                 & 2.9(-6)              & 2.9(-6)              & 3.0(-6)    \\
    \hline
    \end{tabular}
\vspace{0.1 cm}
\end{table}

\begin{table*}[tb]
   \caption{Elemental abundances for T CrB models with M$_{\rm WD} = 1.38$ M$_\odot$.}
    \label{Table13}
    \centering   
    \begin{tabular}{l c c c c c}
    \hline\hline\
    Model                                 & 138A                 & 138B                 & 138C                 & 138D                 & 138E \\                  
    M$_{\rm WD}$(M$_\odot$)               & 1.38                 & 1.38                 & 1.38                 & 1.38                 & 1.38 \\ 
    L$_{\rm WD}$(L$_\odot$)               & 1.0                  & 0.1                  & 0.01                 & 0.1                  & 0.1  \\ 
    Z$_{\rm acc}$                         & 0.014 (Z$_\odot$)    & 0.014 (Z$_\odot$)    & 0.014 (Z$_\odot$)    & 0.0014 (0.1 Z$_\odot$) & 0.14 (10 Z$_\odot$) \\
    \hline
    Z$_{eje}$                             & 0.014                & 0.015                & 0.015                & 0.0015               &  0.14 \\
    X(H)                                  & 6.0(-1)              & 6.0(-1)              & 6.0(-1)              & 5.9(-1)              &  5.9(-1) \\
    X(He)                                 & 3.9(-1)              & 3.9(-1)              & 3.9(-1)              & 4.1(-1)              &  2.8(-1)   \\
    X(Li)                                 & -                    & -                    & -                    & -                    &  2.0(-10)  \\
    X(Be)                                 & -                    & -                    & -                    & -                    &  1.2(-6)   \\
    X(B)                                  & -                    & -                    & -                    & -                    &   -  \\
    X(C)                                  & 3.4(-3)              & 3.0(-3)              & 2.8(-3)              & 2.6(-4)              &  3.9(-2)   \\
    X(N)                                  & 5.9(-3)              & 6.4(-3)              & 6.6(-3)              & 6.9(-4)              &  5.4(-2)   \\
    X(O)                                  & 1.0(-5)              & 7.6(-6)              & 5.4(-6)              & 4.9(-7)              &  2.9(-4)   \\
    X(F)                                  & 2.7(-9)              & 2.9(-9)              & 3.6(-9)              & 5.7(-6)              &  2.0(-8)   \\
    X(Ne)                                 & 5.1(-4)              & 2.5(-4)              & 1.3(-4)              & 3.7(-10)             &  1.6(-2)   \\
    X(Na)                                 & 1.4(-6)              & 9.9(-7)              & 9.9(-7)              & -                    &  3.0(-5)   \\
    X(Mg)                                 & 1.2(-6)              & 6.8(-7)              & 6.4(-7)              & -                    &  4.2(-5)   \\
    X(Al)                                 & 2.6(-6)              & 1.4(-6)              & 9.8(-7)              & -                    &  9.0(-5)    \\
    X(Si)                                 & 6.8(-4)              & 2.1(-4)              & 8.1(-5)              & 1.0(-10)             &  1.7(-2)  \\
    X(P)                                  & 4.9(-5)              & 1.9(-5)              & 1.1(-5)              & -                    &  1.1(-3)   \\
    X(S)                                  & 3.1(-3)              & 3.2(-3)              & 2.4(-3)              & 6.4(-5)              &  5.3(-3)   \\
    X(Cl)                                 & 9.9(-5)              & 1.1(-4)              & 9.6(-5)              & 1.0(-5)              &  7.1(-4)   \\
    X(Ar)                                 & 7.1(-5)              & 8.0(-5)              & 6.4(-5)              & 1.0(-5)              &  3.8(-4)   \\
    X(K)                                  & 1.5(-4)              & 1.9(-4)              & 1.6(-4)              & 2.5(-5)              &  5.2(-5)   \\
    X(Ca)                                 & 4.3(-4)              & 1.3(-3)              & 2.8(-3)              & 5.5(-4)              &  6.6(-4)   \\
    X(Sc)                                 & 2.5(-7)              & 4.2(-7)              & 7.6(-7)              & 5.7(-7)              &  4.7(-7)   \\
    X(Ti)                                 & 2.9(-6)              & 3.0(-6)              & 3.3(-6)              & 3.6(-6)              &  2.9(-5)   \\
    \hline
    \end{tabular}
\vspace{0.1 cm}
\end{table*}

The influence of the initial WD luminosity and the mass-accretion rate on
the elemental abundances of the ejecta can be drawn from a comparison of models
with the same WD mass. Indeed, for models with M$_{\rm WD} = 1.3$ M$_\odot$,
differences in mass fractions by a factor of 3 or larger affect only P and S.
For models with M$_{\rm WD} = 1.35$ M$_\odot$, differences encompass elements such as
Ne, Si, K, Ca, and Sc. For models with M$_{\rm WD} = 1.35$ M$_\odot$, differences
in elemental abundances affect Ne, Si, P, Ca, and Sc\footnote{Multiple 
differences are found when models of different metallicity are considered 
(see Table 13 for details). The largest variations, when comparing models with 
Z/Z$_\odot$ = 0.1, 1 and 10, affect Ne, Si, and P.}. 
A cross-comparison of the variation of elemental abundances between models with 
different
WD masses (for the same initial luminosity) reveals differences
in a large variety of species, including Ne, Na, Mg, Al, Si, P, S, Ar, K, Ca, and Sc. 
Such differences in the elemental abundances of the ejecta could help us 
discriminate between the different T CrB models reported in this work. 
Accordingly, we urge researchers to take early spectroscopic measurements of the forthcoming
outburst of T CrB with the goal of determining the elemental abundances in the ejecta
and, in turn, the conditions under which outbursts in this system take place.

\subsubsection{Gamma-ray emission} 
Previous observations of T CrB (see Sect. 2) have not revealed significant deviations from solar abundances. 
This justifies our choice of solar composition for the plasma transferred from the secondary star, in most of the simulations reported in this work. 
In this context, the amount of radioactive species in the ejecta is expected to be small. 

Figure 1 shows the predicted gamma-ray spectra for a typical case, Model 130C, one
of the models with higher $^{22}$Na (and also $^{18}$F) in the ejecta (see the appendix). The gamma-ray emission has been
computed with a Monte Carlo code, as described in G\'omez-Gomar et al. (1998). 
Spectra at different times after T$_{max}$ are dominated by the 
presence of two prominent lines, at 511 keV (generated by electron-positron annihilation, 
with positrons released in the decay of $^{18}$F and, to some extent, 
$^{13}$N) and 1275 keV (due to $^{22}$Na decay). A lower-energy continuum, powered by 
Comptonization of 511 keV photons --- mainly effective up to 18 hr after T$_{max}$ --- plus contribution 
from positronium emission --- clearly seen at 72 hr, when the ejecta is less dense and Comptonization becomes less effective --- 
is also noticed. However, fluxes corresponding to all these features are orders of magnitude smaller than those 
typically reported for standard models of classical novae with pre-enrichment (see, e.g., Hernanz 2008). 
Accordingly, no detectable gamma-ray lines are expected for the forthcoming outburst of T CrB.

\begin{figure}[tbh]
  \resizebox{\hsize}{!}{\includegraphics{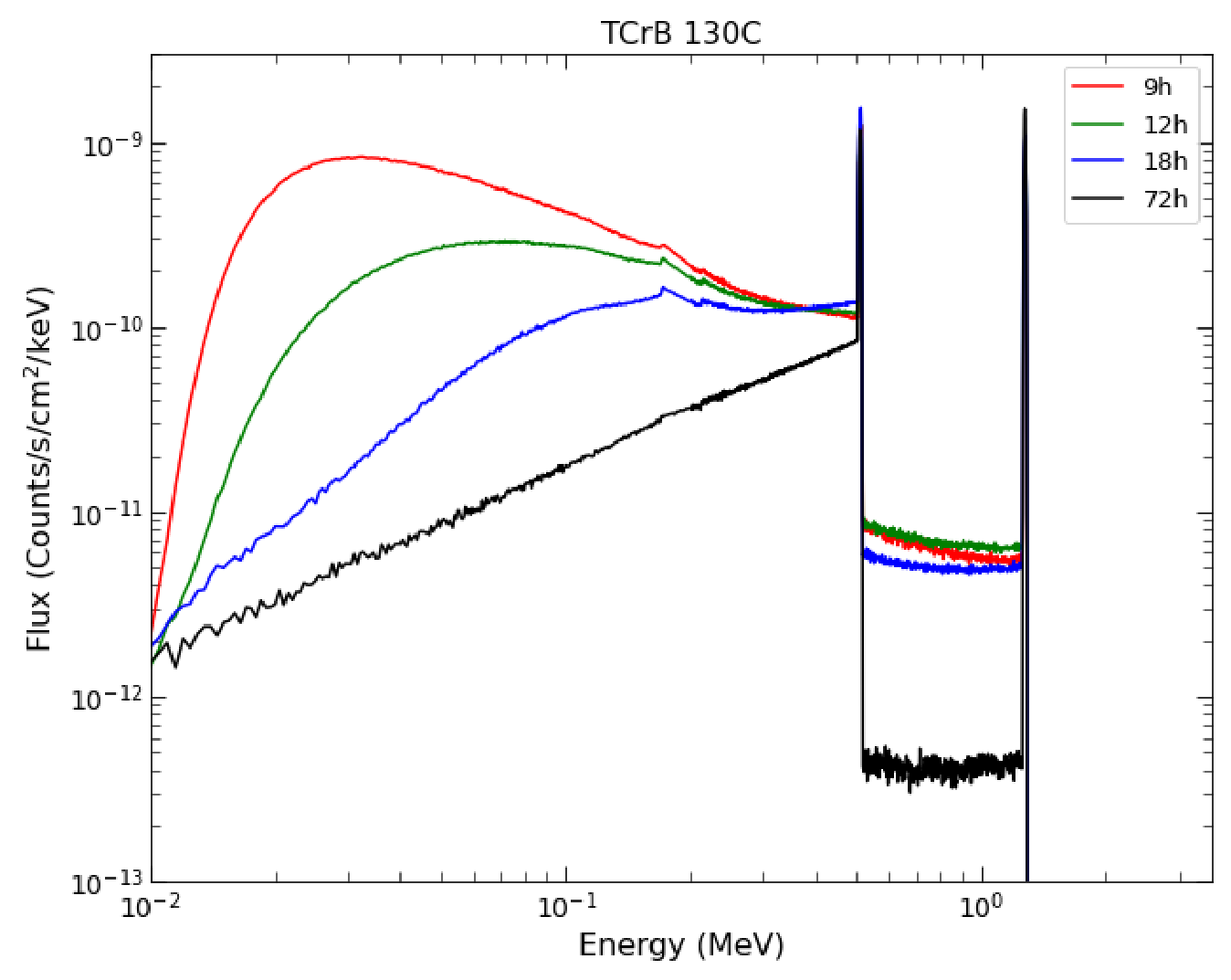}}
  \caption{Early gamma-ray spectra of Model 130C showing the 511-keV line, the lower-energy continuum, and the 1275-keV line, from 9 hr to 3 days after the explosion (T$_{max}$).}
\label{Spectra_1}
\end{figure}

\section{Discussion}
\label{UScoConclusions}

\subsection{Models with mixing}
The limited nuclear activity expected during the outbursts of T CrB
(see Sect. 4) results from the limited mixing between the accreted
material and the outer layers of the WD.
This assumption is supported by the near-solar metallicities typically observed
in the ejecta of most RNe, which contrasts sharply with the
higher metallicities generally inferred in classical nova ejecta, typically
ranging from Z $\sim 0.2 - 0.5$.

Several mixing mechanisms have been proposed in the literature to 
explain the enrichment of the accreted material with WD matter. 
These include diffusion-induced mixing (Prialnik \& Kovetz 1984; Kovetz \& Prialnik 1985; 
Iben et al. 1991, 1992; Fujimoto \& Iben 1992), shear mixing (Durisen 1977; 
Kippenhahn \& Thomas 1978; MacDonald 1983; Livio \& Truran 1987; 
Kutter \& Sparks 1987; Sparks \& Kutter 1987; Bellomo, Shore \& Jos\'e 2024), 
convective overshoot-induced flame propagation (Woosley 1986), mixing via gravity 
wave breaking on the WD surface (Rosner et al. 2001; Alexakis et al. 2004), 
and mixing driven by Kelvin-Helmholtz hydrodynamic instabilities in multidimensional 
nova models (Glasner \& Livne 1995; Glasner et al. 1997, 2005, 2007, 2012; 
Kercek et al. 1998, 1999; Casanova et al. 2010, 2011a, 2011b, 2016, 2018; 
Jos\'e, Shore \& Casanova 2020). It is likely that a combination of these 
mechanisms contributes to mixing during a nova outburst. However, the short time 
interval between outbursts in RNe likely restricts the effectiveness 
of the primary mixing processes responsible for the metallicity enhancement 
typically observed in classical novae.

For completeness, and to evaluate the impact of mixing on the nucleosynthesis 
occurring during RN outbursts, we conducted an additional 
series of simulations. These simulations were performed for WDs 
with masses of 1.35 M$_\odot$ and 1.38 M$_\odot$, assuming different 
(but limited) degrees of mixing between the (solar) accreted envelope and 
matter from the outermost WD layers: 10\% and 25\%. Additionally, 
we considered two distinct compositions for the WD hosting the explosion, 
carbon-oxygen (CO) and oxygen-neon (ONe), as the nature of the compact star in 
RNe remains a subject of debate. 
 The main characteristics of these RN outbursts with mixing are 
summarized in Tables A.4 and A.5. 

The injection of $^{12}$C into the envelope, originating from the underlying substrate, 
significantly alters the explosion dynamics and associated nucleosynthesis. 
A comparison of model 135C with models 135CO1 (10\% mixing) and 135CO2 (25\% mixing), 
which feature a highly C-rich substrate (X($^{12}$C) = 0.495), reveals a shortened 
accretion phase in the mixing models. This results in a net decrease in the total mass 
accreted and a corresponding reduction in the pressure at the base of the envelope at ignition. 
Consequently, the maximum temperature achieved during the explosion and the ejected mass 
are both reduced in these models with mixing.
In contrast, the substantially lower $^{12}$C content in the outer layers of an ONe 
WD (X($^{12}$C) = $9.16 \times 10^{-3}$) minimally impacts the explosion properties 
in models 135ONe1 (10\% mixing) and 135ONe2 (25\% mixing). 
Notably, in model 135C, X($^{12}$C) = $2.33 \times 10^{-3}$ in the accreted envelope. 
However, mixing a solar-composition envelope with an ONe-rich substrate reduces the hydrogen content, 
leading to comparable values of the product X($^{12}$C) $\times$ X($^1$H) in the envelope for 
models 135C, 135ONe1, and 135ONe2.

In terms of associated nucleosynthesis, aside from $^1$H and $^4$He, 
the most abundant species in the ejecta of model 135C are $^{12,13}$C, $^{14,15}$N, 
$^{32}$S, and $^{40}$Ca. In models 135CO1 and 135CO2, $^{12,13}$C and $^{14}$N are 
overproduced by an order of magnitude compared to model 135C. 
Notably, $^{15}$N reaches X($^{15}$N) = 0.11 in model 135CO2, which is nearly 40 times the 
amount ejected in model 135C.
Other elements significantly synthesized in the mixing models include $^{20}$Ne and $^{28}$Si 
in both 135CO1 and 135CO2, as well as $^{16,17}$O in model 135CO2.
Models that incorporate mixing with an ONe-rich substrate are characterized by ejecta enriched 
in $^{14,15}$N, $^{12,13}$C, $^{20}$Ne, $^{28,30}$Si, and $^{32}$S (models 135ONe1 and 135ONe2), 
along with $^{17}$O and $^{31}$P (model 135ONe2).
But the most striking difference between non-mixing and mixing models lies in the 
synthesis of the gamma-ray emitters $^{18}$F, $^{7}$Be, $^{22}$Na, and $^{26}$Al. 
Model 135CO produces negligible amounts of $^{7}$Be (X($^{7}$Be) $\sim 10^{-11}$) 
and low mass fractions of $^{18}$F ($\sim 10^{-9}$, one hour after peak temperature), 
as well as $^{22}$Na and $^{26}$Al ($\sim 10^{-7}$).
In contrast, models with mixing show a substantial increase in the production of these nuclides. 
For models 135CO1 and 135CO2:
X($^{7}$Be) $\sim 10^{-6} - 10^{-5}$,
X($^{18}$F) $\sim 10^{-7} - 10^{-5}$,
X($^{22}$Na) $\sim 10^{-6}$, and
X($^{26}$Al) $\sim 10^{-6} - 10^{-5}$.
Similarly, for models 135ONe1 and 135ONe2:
X($^{7}$Be) $\sim 10^{-8} - 10^{-7}$,
X($^{18}$F) $\sim 10^{-8} - 10^{-6}$, and
X($^{22}$Na) and X($^{26}$Al) $\sim 10^{-5} - 10^{-4}$.
Future space missions with enhanced sensitivities may have the capability to detect 
some of these gamma-ray lines, particularly the 511-keV line and the associated 
lower-energy continuum, provided that enough mixing occurs. Indeed, such observations could provide valuable constraints on the 
extent of mixing in RNe and offer insights into the nature of the 
underlying WD hosting the explosion (CO or ONe)\footnote{Trends similar 
to those observed in the 1.35 M$_\odot$ models are also found in the 1.38 M$_\odot$ models. 
See Tables A.4 and A.5 for details.}.

\subsection{Models with 1.2 M$_\odot$ white dwarfs}
 
Although the most recent estimates for T CrB suggest the presence of a very massive WD,
with M$_{\rm WD} > 1.3$ M$_\odot$ (Zamanov et al. 2003; Hachisu \& Kato 2001; Shara et al. 2018),
the existence of RNe with WD masses around 1.2 M$_\odot$, such as 
T Pyx, IM Nor, and CI Aql (Shara et al. 2018),
stresses the need of characterizing outbursts occurring on the surfaces of less massive WDs.
To this end, we included three additional models involving 1.20 M$_\odot$ WDs 
with initial luminosities of 0.01 L$_\odot$, 0.1 L$_\odot$, and 1 L$_\odot$. These WDs accrete 
solar-composition material at the specific mass-accretion rate that triggers an explosion after 
$80 \pm 0.5$ yr of accretion. The main characteristics of these models are summarized in Table A.6.

One of the most notable differences between these models and those computed for more massive WDs
lies in the amount of mass accreted and ejected. For example, the 1.20 M$_\odot$ models typically 
show values that are 2 to 3 times greater than those found in the 1.30 M$_\odot$ models.
However, the lower surface gravity of a 1.20 M$_\odot$ WD results in lower maximum densities and 
pressures at the base of the accreted envelope. This leads to less violent outbursts, characterized 
by lower peak temperatures and a more moderate nuclear activity. The endpoint of this activity, 
defined as the heaviest isotope whose abundance changes by more than a factor of 2 relative to its 
initial value, is limited to $^{37}$Cl in the 1.20 M$_\odot$ models. In comparison, the endpoints 
reach $^{39}$K and $^{46}$Ti in the 1.30 and 1.38 M$_\odot$ models, respectively.

Nucleosynthesis is essential for distinguishing the mass of the underlying WD hosting the explosion 
in T CrB. While the 1.38 M$_\odot$ models presented in this study exhibit ejecta significantly enriched in 
S, Cl, K, Ca, and Sc relative to solar (i.e., initial) abundances, the 1.20 M$_\odot$ and 1.30 M$_\odot$ models 
generally show solar or slightly subsolar abundances of S, K, Ca, and Sc in the ejecta (except for the 
relatively energetic outburst of model 130C, which yields some enhancement of S and K; see Table A.6 for details).
Moreover, Si is enhanced in the 1.20 M$_\odot$ and 1.30 M$_\odot$ models, whereas its abundance remains 
near-solar or significantly depleted in the 1.38 M$_\odot$ models.
Spectroscopic determinations of the elemental abundances of these key species could prove critical for 
determining the mass of the underlying WD in T CrB.

\section{Conclusions}

The next outburst of T CrB, expected to occur
imminently, motivated us to conduct a thorough examination of the main
characteristics of this system
and the conditions needed to power an outburst with a periodicity
of about 80 yr. To this end, we constructed
11 new hydrodynamic models of the explosion of T CrB for different combinations of parameters (i.e., mass
and initial luminosity of the WD, metallicity, and mass-accretion rate)
that power a nova explosion in approximately 80 yr.
We also reported on 8 additional hydrodynamic models
that include mixing at the interface between the accreted envelope and the outermost layers of the underlying WD, 
and 3 models for 1.20 M$_\odot$ WDs.

As reported in the literature (see, e.g., Starrfield et al. 1988, 
Hernanz \& Jos\'e 2008, Shara et al. 2018) and confirmed by the simulations reported in this work, 
extremely high mass-accretion rates, as well as high WD masses and luminosities, are necessary to drive outbursts with short recurrence periods, such as those observed in RNe. 
Specifically, for a system like T CrB (with $\tau_{\rm rec} \sim 80$ yr), we find that accretion rates of $\dot M_{\rm acc} \sim 10^{-8} - 10^{-7}$ M$_\odot$ yr$^{-1}$ are required for 
WD masses of M$_{\rm WD} \sim 1.30 - 1.38$ M$_\odot$ and luminosities of L$_{\rm WD} \sim 0.01 - 1$ L$_\odot$.
Models with lower initial WD luminosities, less massive WDs, or lower metallicities in the accreted plasma require higher mass-accretion rates to power an explosion within 80 yr.
Models with lower metallicities or initial WD luminosities result in higher accreted masses and ignition pressures. These conditions drive more violent outbursts, characterized by higher 
peak temperatures, higher ejected masses, and greater kinetic energies.

This study\footnote{Another theoretical study focused on T CrB (Starrfield et al. 2025, ApJ 982, 89) has been published while 
this paper was already under review.} has yielded key findings, which are summarized below:

\begin{itemize}
\item  Aside from H and $^4$He, the ejecta are predominantly composed of isotopes from the CNO cycle, including $^{14,15}$N and $^{12,13}$C. In some models, significant amounts of $^{20}$Ne, $^{28}$Si, $^{32}$S, and $^{40}$Ca are also present.

\item Significant variations in the chemical composition are observed within the ejecta of 1.38 M$_\odot$ models with different initial metallicities. For instance, the ejecta in Model 138D (Z = 0.1 Z$_\odot$) show notable contributions from $^{40}$Ca, $^{14,15}$N, $^{12,13}$C, and $^{44}$Ca. In contrast, Model 138E (Z = 10 Z$_\odot$) exhibits a shift in dominance toward $^{14,15}$N, $^{12,13}$C, $^{20}$Ne, and $^{28}$Si.

\item  A comparative analysis of elemental abundance variations across models with different WD masses, but identical initial luminosities, reveals substantial differences in a wide range of species, including Ne, Na, Mg, Al, Si, P, S, Ar, K, Ca, and Sc. These variations within the ejecta provide a potential means to distinguish between the various T CrB models discussed in this work. 
Specifically, while the 1.38 M$_\odot$ models exhibit ejecta significantly enriched in  S, Cl, K, Ca, and Sc relative to solar abundances, the 1.20 M$_\odot$ and 1.30 M$_\odot$ models generally show solar or slightly subsolar abundances of S, K, Ca, and Sc in the ejecta (except for model 130C, which yields some enhancement of S and K). Moreover, Si is enhanced in the 1.20 M$_\odot$ and 1.30 M$_\odot$ models, whereas its abundance remains 
near-solar or significantly depleted in the 1.38 M$_\odot$ models.

\item  Gamma-ray lines (or the continuum) are not expected to be detectable during the forthcoming outburst of T CrB due to the extremely low yields of 
gamma-ray-emitting isotopes in the ejecta, such as $^7$Be, $^{18}$F, $^{22}$Na, and $^{26}$Al. However, test models that incorporate mixing with a CO- or ONe-rich substrate 
--- characteristic of the outer layers of the underlying WD --- reveal significant differences in the predicted emission profiles. 
Future space missions with improved sensitivity might detect the most prominent gamma-ray signatures, if enough mixing occurs. 
Such observations could provide constraints on the degree of mixing in RNe and help us identify the nature of the WD (CO or ONe) that hosts the explosion.

\item  Additionally, we emphasize the importance of conducting early spectroscopic measurements during the next outburst of T CrB aimed at determining the abundances in the ejecta; this would shed light on the physical conditions that govern outbursts in this system.
\end{itemize}

\begin{acknowledgements} 
We thank the anonymous referee for constructive comments, which  helped  improve the quality of the manuscript.
This work has been partially
supported by the Spanish MINECO grants PID2023-148661NB-I00 and PID2023-149918NB-I00, the program Unidad de Excelencia Mar\'{i}a de
Maeztu CEX2020-001058-M, the E.U. FEDER funds, and the AGAUR/Generalitat de Catalunya grants SGR-386/2021 and SGR-1526/2021.
\end{acknowledgements}

\newpage
\newpage

\begin{appendix}
\section{Detailed tables for the T CrB models computed in this work}

\onecolumn

\begin{longtable}{lccc}
\caption{\label{table4full} Properties of T CrB models$^a$ with M$_{\rm WD} = 1.30$ M$_\odot$.}\\
\hline\hline
Model                 & 130A                 & 130B                 & 130C      \\                  
M$_{\rm WD}$(M$_\odot$)               & 1.30                 & 1.30                 & 1.30      \\ 
L$_{\rm WD}$(L$_\odot$)               & 1.0                  & 0.1                  & 0.01      \\ 
Z$_{\rm acc}$                         & 0.014 (Z$_\odot$)    & 0.014 (Z$_\odot$)    & 0.014 (Z$_\odot$)   \\
$\dot {\rm M}_{\rm acc}$ (M$_\odot$ yr$^{-1}$)     &$6.10 \times 10^{-8}$ &$7.16 \times 10^{-8}$ &$1.05 \times 10^{-7}$ \\
\hline
\endfirsthead
\caption{continued.}\\
\hline\hline
           &             &              &     \\                  
\hline
\endhead
\hline
\endfoot
    \hline
    $\Delta$M$_{\rm acc}$($10^{-6}$ M$_\odot$)& 5.40                 & 6.24                 & 8.66  \\
    $\rho_{max}$($10^3$ g cm$^{-3}$)      & 2.16                 & 2.53                 & 3.93  \\
    T$_{max}$($10^8$ K)                   & 2.01                 & 2.08                 & 2.22  \\
    P$_{max}$($10^{19}$ dyn cm$^{-2}$)    & 1.17                 & 1.35                 & 1.88  \\
    K($10^{44}$ ergs)                     & 2.55                 & 2.85                 & 3.91   \\
    v(km s$^{-1}$)                        & 2502                 & 2459                 & 2301   \\
    $\Delta$M$_{eje}$($10^{-6}$ M$_\odot$)& 3.53                 & 4.07                 & 6.29      \\
    Z$_{eje}$                             & 0.014                & 0.014                & 0.014     \\
\end{longtable}

$^a$Mass fractions listed for $> 10^{-10}$.

$^b$Mass fractions correspond to t = 1 hr after T$_{max}$.

\onecolumn

\begin{longtable}{lccc}
\caption{\label{table4full} Properties of T CrB models$^a$ with M$_{\rm WD} = 1.35$ M$_\odot$.}\\
\hline\hline
Model                 & 135A                 & 135B                 & 135C      \\                  
M$_{\rm WD}$(M$_\odot$)               & 1.35                 & 1.35                 & 1.35      \\ 
L$_{\rm WD}$(L$_\odot$)               & 1.0                  & 0.1                  & 0.01      \\ 
Z$_{\rm acc}$                         & 0.014 (Z$_\odot$)    & 0.014 (Z$_\odot$)    & 0.014 (Z$_\odot$)   \\
$\dot {\rm M}_{\rm acc}$ (M$_\odot$ yr$^{-1}$)     &$1.68 \times 10^{-8}$ &$2.07 \times 10^{-8}$ &$3.85 \times 10^{-8}$ \\
    \hline
\endfirsthead
\caption{continued.}\\
\hline\hline
           &             &              &     \\                  
\hline
\endhead
\hline
\endfoot
    \hline
    $\Delta$M$_{\rm acc}$($10^{-6}$ M$_\odot$)& 1.82                 & 2.14                 & 3.49  \\
    $\rho_{max}$($10^3$ g cm$^{-3}$)      & 3.43                 & 4.02                 & 7.01  \\
    T$_{max}$($10^8$ K)                   & 2.36                 & 2.45                 & 2.71  \\
    P$_{max}$($10^{19}$ dyn cm$^{-2}$)    & 1.90                 & 2.23                 & 3.66  \\
    K($10^{44}$ ergs)                     & 1.73                 & 1.99                 & 2.98   \\
    v(km s$^{-1}$)                        & 3558                 & 3531                 & 3361   \\
    $\Delta$M$_{eje}$($10^{-6}$ M$_\odot$)& 1.19                 & 1.40                 & 2.29      \\
    Z$_{eje}$                             & 0.015                & 0.015                & 0.015     \\
\end{longtable}

$^a$Mass fractions listed for $> 10^{-10}$.

$^b$Mass fractions correspond to t = 1 hr after T$_{max}$.

\onecolumn

\begin{longtable}{lccccc}
\caption{\label{table5full} Properties of T CrB models$^a$ with M$_{\rm WD} = 1.38$ M$_\odot$.}\\
\hline\hline
Model                                 & 138A                 & 138B                 & 138C                 & 138D                 & 138E \\                  
M$_{\rm WD}$(M$_\odot$)               & 1.38                 & 1.38                 & 1.38                 & 1.38                 & 1.38 \\ 
L$_{\rm WD}$(L$_\odot$)               & 1.0                  & 0.1                  & 0.01                 & 0.1                  & 0.1  \\ 
Z$_{\rm acc}$                         & 0.014 (Z$_\odot$)    & 0.014 (Z$_\odot$)    & 0.014 (Z$_\odot$)    & 0.0014 (0.1 Z$_\odot$)  & 0.14 (10 Z$_\odot$) \\
$\dot {\rm M}_{\rm acc}$ (M$_\odot$ yr$^{-1}$)     &$7.57 \times 10^{-9}$ &$1.07 \times 10^{-8}$ &$1.67 \times 10^{-8}$ &$2.45 \times 10^{-8}$ &$3.55 \times 10^{-9}$ \\
    \hline
\endfirsthead
\caption{continued.}\\
\hline\hline
           &            &              &             &             &  \\                  
\hline
\endhead
\hline
\endfoot
    \hline
    $\Delta$M$_{\rm acc}$($10^{-6}$ M$_\odot$)&  0.808               & 1.06                 & 1.48                 & 2.62                 & 0.525 \\
    $\rho_{max}$($10^3$ g cm$^{-3}$)      &  4.19                & 5.38                 & 8.12                 & 11.4                 & 3.69  \\
    T$_{max}$($10^8$ K)                   &  2.54                & 2.71                 & 2.92                 & 3.25                 & 2.33  \\
    P$_{max}$($10^{19}$ dyn cm$^{-2}$)    &  2.39                & 3.12                 & 4.40                 & 7.72                 & 1.56  \\
    K($10^{44}$ ergs)                     &  1.17                & 1.44                 & 1.93                 & 2.80                 & 1.03  \\
    v(km s$^{-1}$)                        &  4649                & 4553                 & 4093                 & 3998                 & 4828  \\
    $\Delta$M$_{eje}$($10^{-6}$ M$_\odot$)& 0.474                & 0.619                & 0.972                & 1.54                 &  0.381\\
    Z$_{eje}$                             & 0.014                & 0.015                & 0.015                & 0.0015               &  0.14 \\
\end{longtable}

$^a$Mass fractions listed for $> 10^{-10}$.

$^b$Mass fractions correspond to t = 1 hr after T$_{max}$.

\begin{longtable}{lcccc}
\caption{\label{Table14} T CrB models$^a$ with mixing for M$_{\rm WD} = 1.35$ M$_\odot$.}\\
\hline\hline
Model                 & 135CO1               & 135CO2               & 135ONe1              & 135ONe2 \\                  
M$_{\rm WD}$(M$_\odot$)               & 1.35                 & 1.35                 & 1.35                 & 1.35 \\ 
L$_{\rm WD}$(L$_\odot$)               & 0.01                 & 0.01                 & 0.01                 & 0.01  \\ 
WD Composition                        & CO                   & CO                   & ONe                  & ONe   \\
Pre-enrichment                        & 10\%                 & 25\%                 & 10\%                 & 25\% \\
$\dot {\rm M}_{\rm acc}$ (M$_\odot$ yr$^{-1}$)     &$2.68 \times 10^{-8}$ &$2.25 \times 10^{-8}$ &$4.0 \times 10^{-8}$ &$3.7 \times 10^{-8}$ \\
    \hline
\endfirsthead
\caption{continued.}\\
\hline\hline
          &            &            &            &  \\                  
\hline
\endhead
\hline
\endfoot
    \hline
    $\Delta$M$_{\rm acc}$($10^{-6}$ M$_\odot$)& 2.67                 & 2.32                 & 3.75                 & 3.48 \\
    $\rho_{max}$($10^3$ g cm$^{-3}$)      & 6.55                 & 6.60                 & 8.01                 & 8.33  \\
    T$_{max}$($10^8$ K)                   & 2.58                 & 2.55                 & 2.79                 & 2.79  \\
    P$_{max}$($10^{19}$ dyn cm$^{-2}$)    & 2.80                 & 2.44                 & 3.92                 & 3.67  \\
    K($10^{44}$ ergs)                     & 3.44                 & 4.01                 & 4.05                 & 4.93  \\
    v(km s$^{-1}$)                        & 3684                 & 4375                 & 3332                 & 3863  \\
    $\Delta$M$_{eje}$($10^{-6}$ M$_\odot$)& 2.16                 & 1.88                 & 3.03                 &  2.83\\
    Z$_{eje}$                             & 0.11                 & 0.27                 & 0.12                 &  0.26 \\
\end{longtable}

$^a$Mass fractions listed for $> 10^{-10}$.

$^b$Mass fractions correspond to t = 1 hr after T$_{max}$.

\begin{longtable}{lcccc}
\caption{\label{Table15} T CrB models$^a$ with mixing for M$_{\rm WD} = 1.38$ M$_\odot$.}\\
\hline\hline
Model                 & 138CO1               & 138CO2               & 138ONe1              & 138ONe2 \\                  
M$_{\rm WD}$(M$_\odot$)               & 1.38                 & 1.38                 & 1.38                 & 1.38 \\ 
L$_{\rm WD}$(L$_\odot$)               & 0.01                 & 0.01                 & 0.01                 & 0.01  \\ 
WD Composition                        & CO                   & CO                   & ONe                  & ONe  \\
Pre-enrichment                        & 10\%                 & 25\%                 & 10\%                 & 25\% \\
$\dot {\rm M}_{\rm acc}$ (M$_\odot$ yr$^{-1}$)     &$1.15 \times 10^{-8}$ &$0.97 \times 10^{-8}$ &$1.73 \times 10^{-8}$ &$1.62 \times 10^{-8}$ \\
    \hline
\endfirsthead
\caption{continued.}\\
\hline\hline
          &            &            &            &  \\                  
\hline
\endhead
\hline
\endfoot
    $\Delta$M$_{\rm acc}$($10^{-6}$ M$_\odot$)& 1.14                 & 0.991                & 1.59                 & 1.51  \\
    $\rho_{max}$($10^3$ g cm$^{-3}$)      & 7.63                 & 7.64                 & 9.18                 & 9.74  \\
    T$_{max}$($10^8$ K)                   & 2.78                 & 2.74                 & 3.01                 & 3.01  \\
    P$_{max}$($10^{19}$ dyn cm$^{-2}$)    & 3.38                 & 2.95                 & 4.71                 & 4.48  \\
    K($10^{44}$ ergs)                     & 2.23                 & 2.54                 & 2.61                 & 3.20  \\
    v(km s$^{-1}$)                        & 4517                 & 5187                 & 4044                 & 4707  \\
    $\Delta$M$_{eje}$($10^{-6}$ M$_\odot$)& 0.920                & 0.801                & 1.29                & 1.22\\
    Z$_{eje}$                             & 0.12                 & 0.27                 & 0.13                 &  0.27 \\
\end{longtable}

$^a$Mass fractions listed for $> 10^{-10}$.

$^b$Mass fractions correspond to t = 1 hr after T$_{max}$.

\begin{longtable}{lccc}
\caption{\label{Table16} T CrB models$^a$ with M$_{\rm WD} = 1.20$ M$_\odot$.}\\
\hline\hline
Model                 & 120A                 & 120B                 & 120C      \\                  
M$_{\rm WD}$(M$_\odot$)               & 1.20                 & 1.20                 & 1.20      \\ 
L$_{\rm WD}$(L$_\odot$)               & 1.0                  & 0.1                  & 0.01      \\ 
Z$_{\rm acc}$                         & 0.014 (Z$_\odot$)    & 0.014 (Z$_\odot$)    & 0.014 (Z$_\odot$)   \\
$\dot {\rm M}_{\rm acc}$ (M$_\odot$ yr$^{-1}$)     &$1.15 \times 10^{-7}$ &$1.60 \times 10^{-7}$ &$2.80 \times 10^{-7}$ \\
    \hline
\endfirsthead
\caption{continued.}\\
\hline\hline
           &             &              &     \\                  
\hline
\endhead
\hline
\endfoot
    \hline
    $\Delta$M$_{\rm acc}$($10^{-6}$ M$_\odot$)& 11.1                 & 13.9                 & 24.1  \\
    $\rho_{max}$($10^3$ g cm$^{-3}$)      & 1.52                 & 2.12                 & 3.82  \\
    T$_{max}$($10^8$ K)                   & 1.78                 & 1.86                 & 2.09  \\
    P$_{max}$($10^{19}$ dyn cm$^{-2}$)    & 0.814                & 1.03                 & 1.79  \\
    K($10^{44}$ ergs)                     & 2.86                 & 3.44                 & 5.96   \\
    v(km s$^{-1}$)                        & 1842                 & 1799                 & 1567   \\
    $\Delta$M$_{eje}$($10^{-6}$ M$_\odot$)& 7.22                 & 9.11                 & 19.7      \\
    Z$_{eje}$                             & 0.014                & 0.014                & 0.014     \\
\end{longtable}

$^a$Mass fractions listed for $> 10^{-10}$.

$^b$Mass fractions correspond to t = 1 hr after T$_{max}$.

\end{appendix}
\end{document}